\documentclass[12pt]{iopart}

\usepackage{iopams}
\usepackage{graphicx}
\usepackage{amssymb}
\usepackage{threeparttable}
\usepackage{dcolumn}
\usepackage{bm}
\usepackage{cite}
\begin{document}

\title[Probe of deformed halo and 2n-radioactivity at Mg neutron drip-line]{Collapse of N$=$28 magicity in exotic $^{40}$Mg - Probe of deformed halo and 2n-radioactivity at Mg neutron drip-line}
\author{G. Saxena$^{1,\S}$, M. Kumawat$^{2,3}$, R. Sharma$^{3}$ and Mamta Aggarwal$^{4,\dagger}$}
\address{$^{1}$Department of Physics (H\&S), Govt. Women Engineering College, Ajmer-305002, India}
\address{$^{2}$Department of Science and Technology, Faculty of Education and Methodology, Jayoti Vidyapeeth Women's University, Jaipur-303007, India}
\address{$^{3}$Department of Physics, Manipal University Jaipur, Jaipur-303007, India}
\address{$^{4}$Department of Physics, University of Mumbai, Kalina, Mumbai-400098, India.}
\ead{$^{\dagger}$mamta.a4@gmail.com, $^{\S}$gauravphy@gmail.com}
\vspace{10pt}
\begin{indented}
\item[]March 2021
\end{indented}

\begin{abstract}
The exotic phenomenon of two-neutron halos and 2n-radioactivity are explored in the neutron-rich $^{40,42,44}$Mg by employing various variants of the relativistic mean-field approach. The extended tail of spatial density distributions including the enhanced neutron radii and skin thickness, pairing correlations, single-particle spectrum and wave functions predict $^{40,42,44}$Mg to be strong candidates for deformed neutron halos. Weakening of magicity at N$=$28 plays a significant role in the existence of a weakly bound halo in $^{40}$Mg which is currently the heaviest isotope of Mg accessible experimentally. Large deformation, mixing of f-p shell Nilsson orbitals and the valence neutron occupancy of p-states leads to a reduced centrifugal barrier and broader spatial density distributions that favour 2n-radioactivity  in $^{42,44}$Mg.\par
\end{abstract}

\submitto{\JPG}

\section{Introduction}
The nucleus $^{40}$Mg, with magic neutron number N$=$28, is the most neutron-rich isotope of Mg accessible experimentally \cite{baumann2007}. The experimental observation of $\gamma$-ray transitions in $^{40}$Mg \cite{crawford2019} indicate that the structure of $^{40}$Mg is different from the neighbouring isotopes $^{36,38}$Mg in contrast to the theoretical expectations \cite{crawford2019}. This could possibly be due to two weakly bound valence neutrons at the Fermi surface that may couple to the classically forbidden continuum where the nucleon binding is dominated by pairing correlations rather than the mean-field of core to form a halo \cite{tanihata}. Weak binding along with the extended spatial density distributions, the coupling of discrete bound states to the continuum, possible changes in the magicity and the phenomenon of 2n-radioactivity in $^{40}$Mg and its neighbouring isotopes, need to be investigated to understand the exotic nature of this region.\par

The 2n-halo systems may exhibit strong two-neutron correlations \cite{zhukov1993}. A recent 2n-decay study \cite{MONA} suggests that the 2n-unbound systems may provide a unique opportunity to measure the correlations of the neutrons and the possibility of 2n-radioactivity \cite{MONA,leblond2018}. So far the ground state 2n-decay has been reported for $^{16}$Be \cite{spyrou} and $^{26}$O \cite{kohley,grigorenko1} whereas the heavier nuclei are difficult to access and thus remain unexplored. Various theoretical calculations have indicated that the neutron-rich nuclei $^{42,44}$Mg are 2n-unbound due to the two-neutron separation energy (S$_{2n}$) $<$0 and one neutron separation energy (S$_{1n}$)$>$0 and predicted \cite{LLI,zhou,meng2015} to have deformed 2n-halo which points to the role of dineutron correlations of the valence neutrons in 2n-radioactivity. On the other hand, $^{40}$Mg being relatively weakly bound as per the mass surface extrapolation of AME2020 \cite{ame2020} (S$_{2n}$ $=$ 0.67$\pm$0.71 MeV), does not indicate 2n-radioactivity. The occupancy of the 2p$_{3/2}$ neutron orbital and the fact that the neighbouring odd-A $^{39}$Mg is unbound, point towards pairing correlations and the possibility of a 2n-halo in $^{40}$Mg.\par

The enhancement of matter radii, which is a signature of nuclear halos, has been reported theoretically recently for $^{40}$Mg \cite{watanabe2014} due to large deformation. This is indicative of the weakening of magicity of N$=$28 \cite{bastin2007,doornebal2013} near the neutron drip-line. Since the neutron-rich N$=$28 isotones play an important role in the nucleosynthesis of heavy Ca-Ti-Cr isotopes, it is crucial to know if the magicity of N$=$28 persists or breaks in the neutron-rich region. Recent mass measurement and spectroscopic experiments have shown that N$=$28 shell closure is weakened in Mg \cite{watanabe2014,crawford2019}, Si \cite{bastin2007,camp2006} and S \cite{gaud2009}. Disappearance of conventional magic numbers like N$=$8 \cite{navin2000,iwasaki2000}, 20 \cite{mueller1984,moto1995}, and 28 \cite{sorlin1993,glasmacher1997,sarazin2000,bastin2007}, and  the emergence of new magic numbers like N$=$16 \cite{ozawa2000} and 34 \cite{step2013} near the drip-lines has been a major breakthrough in the search of exotic features which also poses a challenge to the existing theoretical models to study exotic nuclei. The deformation effects increase the number of particles in the classically forbidden region below the continuum threshold as recently reported for $^{42,44}$Mg \cite{zhang2019}. In view of the possible significance of deformation, pairing correlations, and the collapse of magicity at N$=$28 \cite{watanabe2014,crawford2019} in the existence of neutron halo and 2n-decay in neutron-rich Mg isotopes, we have undertaken a detailed investigation. Mapping of the precise position of Mg drip-line \cite{meng2015}, unveiling of the structural dynamics of $^{40,42,44}$Mg and probing the possible correlation of the halo with 2n-decay, is the objective of the present work.\par

\section{Theoretical Formulation}

We use relativistic mean-field approach (RMF) \cite{Lalazissis09,sugaTMA,pk1,Todd-Rutel05,Chen15,Lalazissis05,Niksic08,Saxenaplb2018,Saxena2017,saxenajpg} for the present study where the various parameterizations/variants of the RMF models have been used for the calculations. The first variant of RMF comprises of the following model Lagrangian density with nonlinear terms for the ${\sigma}$ and ${\omega}$ mesons \cite{Saxenaplb2018,Singh2013,Gambhir1989,Geng2003};

\begin{eqnarray}
       {\cal L}& = &{\bar\psi} [\imath \gamma^{\mu}\partial_{\mu}
                  - M]\psi\nonumber\\
                  &&+ \frac{1}{2}\, \partial_{\mu}\sigma\partial^{\mu}\sigma
                - \frac{1}{2}m_{\sigma}^{2}\sigma^2- \frac{1}{3}g_{2}\sigma
                 ^{3} - \frac{1}{4}g_{3}\sigma^{4} -g_{\sigma}
                 {\bar\psi}  \sigma  \psi\nonumber\\
                &&-\frac{1}{4}H_{\mu \nu}H^{\mu \nu} + \frac{1}{2}m_{\omega}
                   ^{2}\omega_{\mu}\omega^{\mu} + \frac{1}{4} c_{3}
                  (\omega_{\mu} \omega^{\mu})^{2}
                   - g_{\omega}{\bar\psi} \gamma^{\mu}\psi
                  \omega_{\mu}\nonumber\\
               &&-\frac{1}{4}G_{\mu \nu}^{a}G^{a\mu \nu}
                  + \frac{1}{2}m_{\rho}
                  ^{2}\rho_{\mu}^{a}\rho^{a\mu}
                   - g_{\rho}{\bar\psi} \gamma_{\mu}\tau^{a}\psi
                  \rho^{\mu a}\nonumber\nonumber\\
                &&-\frac{1}{4}F_{\mu \nu}F^{\mu \nu}
                  - e{\bar\psi} \gamma_{\mu} \frac{(1-\tau_{3})}
                  {2} A^{\mu} \psi\,\,,
\end{eqnarray}
where $H$, $G$ and $F$ being the field tensors for the vector fields are
defined by
\begin{eqnarray}
                 H_{\mu \nu} &=& \partial_{\mu} \omega_{\nu} -
                       \partial_{\nu} \omega_{\mu}\nonumber\\
                 G_{\mu \nu}^{a} &=& \partial_{\mu} \rho_{\nu}^{a} -
                       \partial_{\nu} \rho_{\mu}^{a}
                     -2 g_{\rho}\,\epsilon^{abc} \rho_{\mu}^{b}
                    \rho_{\nu}^{c} \nonumber\\
                  F_{\mu \nu} &=& \partial_{\mu} A_{\nu} -
                       \partial_{\nu} A_{\mu}\,\,,\nonumber\
\end{eqnarray}
Other symbols have their usual meaning and the details can be found in Refs.$~$ \cite{Boguta77,Boguta83,Furnstahl97}. We have used NL3* parametrization \cite{Lalazissis09} of the RMF model for most of the calculations in this work. It includes linear terms for the $\sigma$, $\omega$ and $\rho$ mesons and the non-linear term for the self interaction of the $\sigma$ meson only. We have also performed few calculations using TMA \cite{sugaTMA} and PK1 \cite{pk1} parameterizations which include nonlinear terms for the ${\omega}$ meson in addition to the non-linear terms of $\sigma$ meson as described in detail in Refs.~\cite{sugaTMA,pk1}. Few other parameters containing non-linear self interaction of $\omega$ meson and the mixed interaction terms for $\omega$ and $\rho$ mesons are FSU-Gold and FSU-Garnet \cite{Todd-Rutel05,Chen15} that have been used in the calculations. \par

The second variant of the RMF model belongs to the category of density dependence of coupling constants for the meson exchange i.e. DD-ME2 and DD-PC1 \cite{Lalazissis05,Niksic08}. The density-dependent meson-exchange model (DD-ME) \cite{Lalazissis05} interaction part of the Lagrangian does not contain any non-linear term, but, the meson-nucleon strengths $g_{\sigma}$, $g_{\omega}$ and $g_{\rho}$ have an explicit density
dependence in the following form:

\begin{equation} \label{eq:Equation1} 
g_{i}(\rho) =  g_{i}(\rho_{sat})f_{i}(x), \,\,\,\,\,\,\, for \,\,\, i = \sigma, \omega 
\end{equation}

where the density dependence is given by
\begin{equation} \label{eq:Equation2} 
f_{i}(x) =  a_{i} \frac {1+b_{i}(x+d_{i})^{2}}{1+c_{i}(x+d_{i})^{2}} 
\end{equation}
in which $x$ is given by $x = \rho/\rho_{sat}$, and $\rho_{sat}$ denotes the baryon density at saturation in symmetric nuclear matter.
For the $\rho$ meson, density dependence is of exponential form and given by
\begin{equation} \label{eq:Equation3} 
f_{\rho}(x) =  exp(-a_{\rho}(x-1)) 
\end{equation}

The effective lagrangian for DD-PC1 model \cite{Niksic08} is analogous to DD-ME2 model but, it does not include the derivative term for mesonic fields and hence they are directly expressed in terms of nucleonic field;

\begin{eqnarray} \label{eq:Lagrangian3} 
{\cal L}_{\it int}=&& -\frac{1}{2}\alpha_{S}(\rho)(\overline{\psi}\psi)(\overline{\psi}\psi) -
\frac{1}{2}\alpha_{V}(\rho)(\overline{\psi}\gamma^{\mu}\psi)(\overline{\psi}\gamma_{\mu}\psi)\nonumber
\\ &&- \frac{1}{2}\alpha_{TV}(\rho)(\overline{\psi}\overrightarrow{\tau}\gamma^{\mu}\psi)(\overline{\psi}\overrightarrow{\tau}\gamma_{\mu}\psi) - \frac{1}{2}\delta_{S}(\rho)(\overline{\psi}\psi)\square(\overline{\psi}\psi) \nonumber 
\end{eqnarray}

In analogy with meson-exchange model (DD-ME) described above, this model contains isoscalar-scalar (S), isoscalar-vector (V) and isovector-vector
(TV) interactions. The coupling constants $\alpha_{i}(\rho)$ are density dependent and have the form \cite{Niksic08}:

\begin{equation} \label{eq:Equation4} 
\alpha_{i}(\rho)  =  a_{i} + (b_{i} + c_{i}x)e^{-d_{i}x}, for \,\,\, i = S, V, TV
\end{equation}

There are three parts to the present approach (i) Determining the precise position of the two-neutron Mg drip-line (ii) Evidences and implications of the weakening of N$=$28 shell closure in exotic $^{40}$Mg. (iii)  Existence of deformed halo in neutron-rich Mg isotopes $^{40,42,44}$Mg and exploring its possible correlation with 2n-decay.

\section{Results and Discussion}
\subsection{Two-Neutron drip-line for Mg}
To probe 2n-radioactivity, which is essentially the energetically allowed simultaneous emission of two-neutrons, we first estimate neutron separation energies S$_{n}$ and S$_{2n}$ for $^{40,42,44}$Mg, shown in Table \ref{s2n-sn}. We use the relativistic mean-field theory with NL3*\cite{Lalazissis09}, TMA\cite{sugaTMA}, PK1 \cite{pk1}, FSUGold\cite{Todd-Rutel05}, FSUGarnet \cite{Chen15}, DDME2 \cite{Lalazissis05} and DDPC1 \cite{Niksic08} for the calculations. We also calculate S$_{n}$ and S$_{2n}$ values by employing the shell-model approach \cite{brown2014} and the macroscopic-microscopic approach using the triaxially deformed Nilsson Strutinsky method (NSM) \cite{Aggarwal2010,Aggarwal2014}. These values are compared with the latest atomic mass evaluation AME2020 \cite{ame2020} and the other theories viz. HFB \cite{goriely}, RCHB \cite{RCHB}, FRDM \cite{frdm2012} and INM \cite{INM}. Since the predictions on the last bound nucleus among the neutron-rich heavy isotopes $^{40,42,44}$Mg are different in different works, the precise position of neutron drip-line for Mg isotopes is still uncertain. Most of the theoretical models as well as the mass surface extrapolation \cite{ame2020} predict $^{40}$Mg to be weakly 2n-bound except for RMF(PK1), FSU-Gold and INM. However, most of the theories suggest $^{42}$Mg  and $^{44}$Mg to be 2n-unbound as seen in Table \ref{s2n-sn}. \par

\begin{table}[htb]
\caption{Neutron separation energies S$_{n}$ and S$_{2n}$ using NL3* \cite{Lalazissis09}, TMA \cite{sugaTMA}, PK1 \cite{pk1}, FSU-Gold \cite{Todd-Rutel05}, FSU-Garnet \cite{Chen15}, DD-ME2 \cite{Lalazissis05}, DD-PC1 \cite{Niksic08}, our calculation with shell-model \cite{brown2014} and NSM \cite{Aggarwal2010,Aggarwal2014}, latest atomic mass evaluation AME2020 \cite{ame2020} and other theories HFB \cite{goriely}, RCHB \cite{RCHB}, FRDM \cite{frdm2012} and INM \cite{INM}.}
\centering
\begin{threeparttable}[b]
\resizebox{0.6\textwidth}{!}{%
{\begin{tabular}{l|r|r|r|r|r|r}
 \hline
 \hline
 \multicolumn{1}{c|}{}&
 \multicolumn{2}{c|}{$^{40}$Mg}&
 \multicolumn{2}{c|}{$^{42}$Mg}&
 \multicolumn{2}{c}{$^{44}$Mg}\\
\cline{2-7}
\multicolumn{1}{c|}{}&
 \multicolumn{1}{c|}{S$_{n}$ }&
 \multicolumn{1}{c|}{S$_{2n}$ }&
 \multicolumn{1}{c|}{S$_{n}$ }&
\multicolumn{1}{c|}{S$_{2n}$ }&
\multicolumn{1}{c|}{S$_{n}$ }&
 \multicolumn{1}{c}{S$_{2n}$ } \\
 \multicolumn{1}{c|}{}&
 \multicolumn{1}{c|}{(MeV)}&
 \multicolumn{1}{c|}{ (MeV)}&
 \multicolumn{1}{c|}{ (MeV)}&
\multicolumn{1}{c|}{ (MeV)}&
\multicolumn{1}{c|}{ (MeV)}&
 \multicolumn{1}{c}{ (MeV)} \\
 \hline
 \hline
  AME2020      & 1.30$\pm$0.72 $\tnote{1}$    &0.67$\pm$0.71 $\tnote{1}$  &  -      & -      &-&-\\
  \hline
  NL3*  & 1.69    &2.45   & 0.24   & -0.06 &0.38&-0.27\\
  TMA   & 1.33    &1.75    & -0.44  & -1.26 &-1.02&-1.19 \\
  PK1   & -0.07   &-0.61   & 0.05   & -0.82 &0.05&-0.82\\
  FSU-Gold    & 2.03    &-1.45   &-0.41   & -1.03 &-0.38&1.43\\
  FSU-Garnet  & 0.28    &0.13    &-1.21   & -2.46 &-1.06&-2.03 \\
  DD-ME2     & 0.28    &0.93    & -0.76  & -1.07 &-0.94&-1.80 \\
  DD-PC1     &0.91     &1.89    & -0.72  &-1.36  &-0.95&-1.86 \\
  \hline
  Shell-Model& 1.38    &2.01    &0.90    &-0.85  &0.91 & -0.42\\
  NSM        & 2.31    &0.10    & 1.22   & -1.66 &0.13&-3.73 \\
  \hline
  HFB        & 1.36    &0.76    & 1.97   & 0.41  &0.39&-1.31 \\
  RCHB       & 2.84    &2.72    & 2.65   & 2.45  &2.20&2.17 \\
  FRDM       & 2.87    &2.83    & 0.92   & -2.23 &0.01&-2.72 \\
  INM        & 0.29    &-0.55   & -0.48  & -5.37 &-&- \\
\hline
   \hline
\end{tabular}}}
\begin{tablenotes}
    \item[1] Estimated using mass surface extrapolation
  \end{tablenotes}
 \end{threeparttable}
\label{s2n-sn}
\end{table}

Table I shows the extremely weak binding in $^{40}$Mg with a small value of positive separation energy, whereas in the case of $^{42,44}$Mg, S$_{2n}$ $<$ 0 and S$_{n}$ $>$ 0 by most of the theories, which means the emission of two valence neutrons or dineutron is allowed but the sequential emission of one neutron is energetically forbidden indicating 2n-radioactivity. A similar process of two proton-radioactivity which occurs when S$_{p}$ $>$ 0 and S$_{2p}$ $<$ 0, was studied in our recent work \cite{Saxena2017} where the existence of 2p-halos in 2p-emitters and the correlation between the 2p-halo and 2p-emission was shown. However, being close to N$=$Z line, the proton drip-line is experimentally easily accessible than the neutron drip-line which is why the 2n-emission is a less explored phenomena and reliable theoretical models are desirable. Since the nuclear halos are expected to have extremely weak binding, as is the case of $^{40}$Mg, we explore the structural aspects for the possible existence of halo in $^{40}$Mg.  \par

\subsection{Weakening of N = 28 shell closure}

\begin{figure}[htb]
\centering
\includegraphics[width=0.6\textwidth]{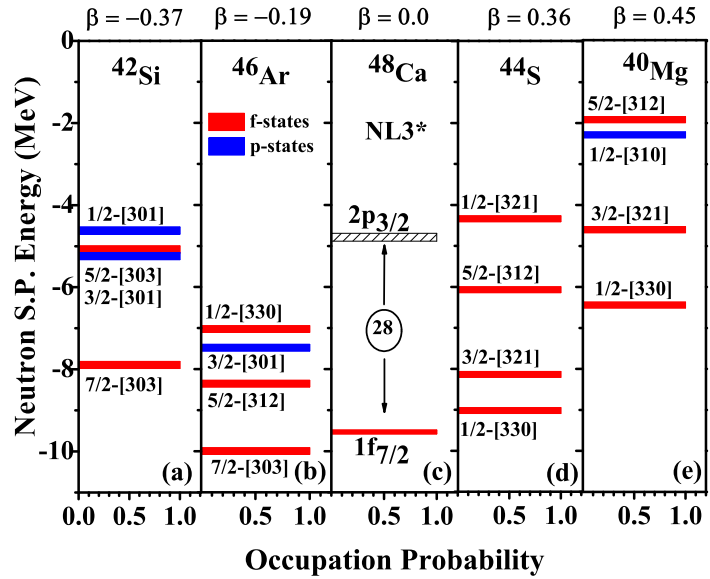}
\caption{(Colour online) Neutron s.p energies and occupancies for N$=$28 isotones using RMF(NL3*) approach.}
\label{fig1}
\end{figure}

There have been predictions of significant deformation in neutron-rich Mg isotopes \cite{nakada2018,Hamamoto} including $^{40}$Mg which has magic neutrons N$=$28. Since a magic nucleus with N$=$28 is typically expected to have a completely filled $\nu$1$f_{7/2}$ shell followed by a large shell gap and zero deformation, $^{40}$Mg does not fit into the criteria of a magic nucleus because of it's large deformation \cite{watanabe2014} (prolate with $\beta$$=$0.453) and partial occupancy in $\nu$1$f_{7/2}$ and $\nu$2$p_{3/2,1/2}$ \cite{crawford2019} states which shows weakening  of N$=$28 magicity \cite{sorlin1993,glasmacher1997,sarazin2000,bastin2007,watanabe2014,crawford2019}. Since the deformation and the  occupancy at Fermi level are correlated, we investigate the neutron single-particle (s.p.) energies along with the deformation and occupancies of $\nu$1$f_{7/2}$ and $\nu$2$p_{3/2}$ states shown in Fig. \ref{fig1} for N$=$28 isotones of Z$=$12 to 20 calculated by using RMF(NL3*) approach. The deformation of these isotones is mentioned at the top of the figure. \par

In Fig. \ref{fig1}, the transition from the zero deformation in stable nucleus $^{48}$Ca to large deformation in neutron-rich isotones (Z$=$12-18) of N$=$28 shows the weakening of magicity of N=28 at neutron drip-line. A large shell gap of around $~$4.7 MeV between $\nu$1$f_{7/2}$ and $\nu$2$p_{3/2}$ with 1.0 occupancy in  $\nu$1$f_{7/2}$ and zero in $\nu$2$p_{3/2}$ (shown by shaded bar) defines the doubly magic nucleus $^{48}$Ca as expected. In well deformed neutron-rich isotones $^{40}$Mg, $^{42}$Si, $^{44}$S and $^{46}$Ar, the mixing of f-p shell Nilsson orbital and the disappearance of the shell gap between $\nu$1$f_{7/2}$ and $\nu$2$p_{3/2}$, in accord with Refs.$~$ \cite{sorlin2008,li2011,afanasjev2015,caurier2004} indicates that the shell closure at N$=$28 no longer exists. In $^{40}$Mg, 6 particles occupy Nilsson f-orbital 5/2-[312], 3/2-[321], 1/2-[330] and the remaining 2 valance particles occupy p-state 1/2-[310] that spend at least some time there, resulting in the smaller centrifugal barrier due to lower orbital angular momentum ($\ell$$=$1) and pairing interactions that may lead to dineutron correlations and contribute to a weakly bound state or a 2n-halo, which enhances the probability of direct 2n-decay.\par

\begin{figure}[htb]
\centering
\includegraphics[width=0.6\textwidth]{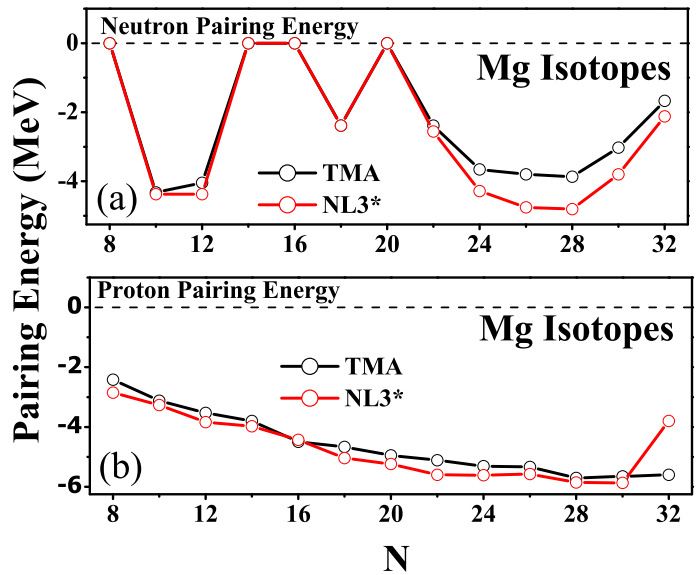}
\caption{(Colour online) Neutron and proton pairing energy contribution for Mg isotopes.}
\label{fig2}
\end{figure}

Fig. \ref{fig2} shows pairing correlations in Mg isotopes from neutron deficient N$=$8 to neutron-rich N$=$32. Non-zero pairing energy  at N$=$28 with the maximum value among other isotopes is a strong evidence of the breaking of magicity at N$=$28. On the other hand, zero neutron pairing energy at N$=$14 points towards the magicity of N$=$14 as seen in  Refs. \cite{thirolf,stanoiu,becheva} and our work \cite{Saxena2017hzo} which indicated towards the weakening of magicity in conventional magic numbers and the emergence of new magic numbers  N$=$14 \cite{mcormick} and 40 \cite{wang1} at the neutron drip-line. However, non-zero pairing interaction in $^{40}$Mg and $^{42,44}$Mg indicates the possibility of halo formation. \par

\begin{figure}[htb]
\centering
\includegraphics[width=0.6\textwidth]{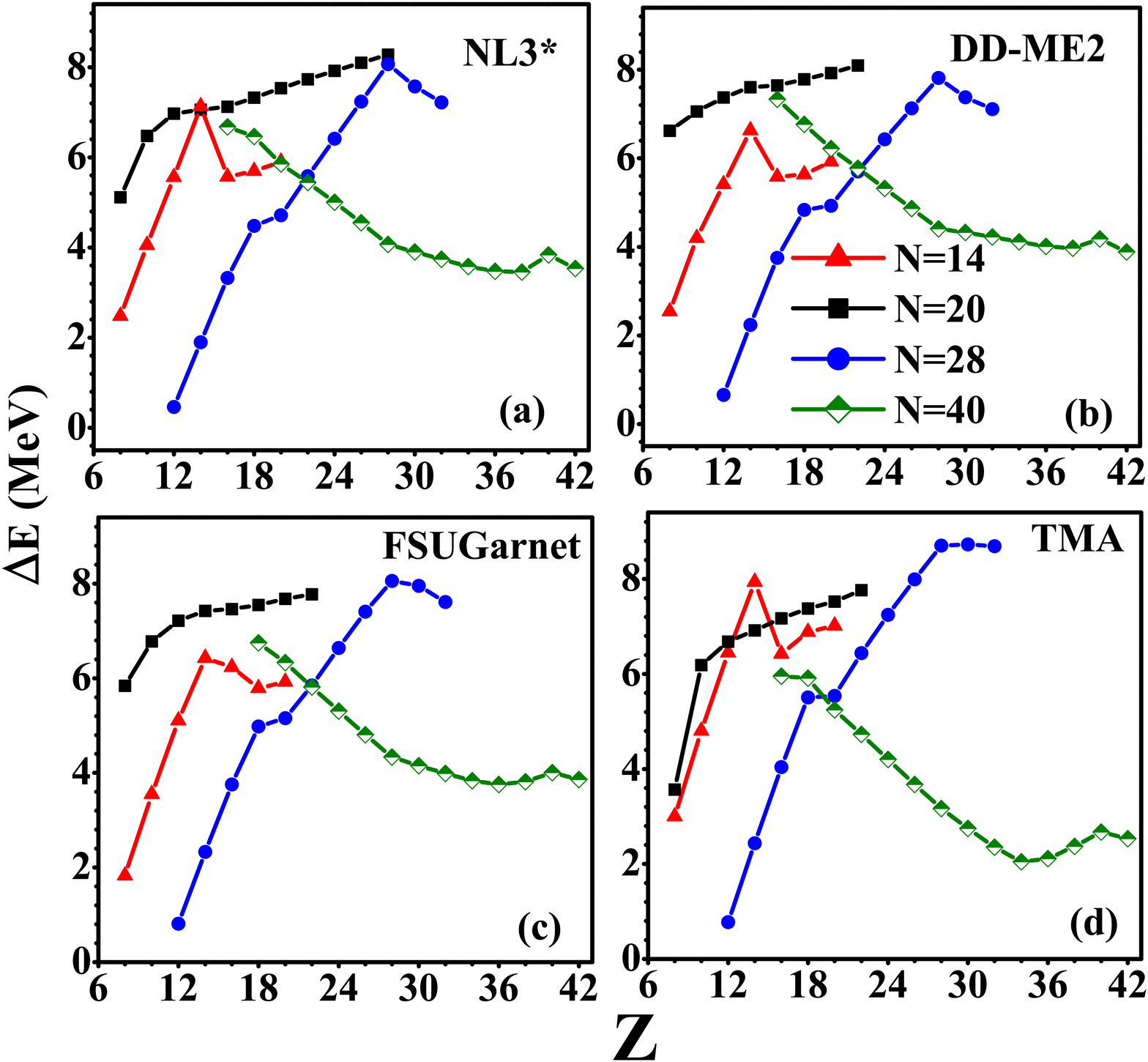}
\caption{(Colour online) Energy shell gap ($\Delta$E) between last filled and first unfilled states of N$=$14, 20, 28 and 40 isotones.}
\label{fig3}
\end{figure}

In view of the changing magicity at neutron drip-line,  we plot and compare the spherical shell gap in terms of the energy difference ($\Delta$E) between the last filled and the first unfilled state of N$=$14, 20, 28 and 40 isotones in Fig. \ref{fig3} in a very novel fashion. We use the parameter sets NL3* \cite{Lalazissis09}, TMA \cite{sugaTMA}, FSU-Garnet \cite{Chen15} and DD-ME2 \cite{Lalazissis05} for the calculation of $\Delta$E.  \par

Fig. \ref{fig3} shows $\Delta$E for the isotonic chains of N$=$14, 20, 28 and 40. Since the shell closures are identified by a large shell gap ($\Delta$E) typically of the order of 6-8 MeV as is seen in conventional magic number N=20 isotones by all the RMF parameters (in Fig \ref{fig3} (a)-(d)). But in case of N$=$28, the shell gap ($\Delta$E) is large around 7-8 MeV showing strong magicity in N$=$28 isotones of nuclei with Z$\ge$ 28, but a steep decline in $\Delta$E between $\nu$1$f_{7/2}$ and $\nu$2$p_{3/2}$ for neutron-rich (Z $<$ 28) isotones strongly supports the weakening of shell closure at N$=$28 in neutron-rich region. The comparison of the shell gap in neutron-rich N$=$14, 20, 28 and 40 isotones shows the smallest value of $\Delta$E for N$=$28 isotones which is even smaller than the non-conventional magic numbers N=14 and 40 isotones, validating the weakening of magicity of N$=$28 at neutron drip-line. The shell gap in $^{40}$Mg is around 1 MeV or less, the smallest of all, confirming the complete collapse of magicity of N$=$28 in $^{40}$Mg.\par
A very recent work by  Reinhard \textit{et al.} \cite{reinhard2020} has proposed that the fourth radial moment (r$_{4}$) can be directly related to the surface thickness of nuclear density. This may be very useful to identify magicity because the asymptotic density values for a magic nucleus are believed to fall off rapidly as compared to the non-magic nucleus. To identify the magicity of N$=$28 in $^{40}$Mg, we evaluate the rms radius (second moment r$_{2}$) and fourth radial moment (r$_{4}$) where the radial moments are defined as \cite{reinhard2020}:

\begin{equation}
r_n \equiv \sqrt[n]{\left\langle r^{n} \right\rangle}\,=\,\left(\frac{\int d^{3}r \, r^n \, \rho(r)}{\int d^{3}r \, \rho(r)}\right)^{1/n} \nonumber
\end{equation}

\begin{figure}[htb]
\centering
\includegraphics[width=0.6\textwidth]{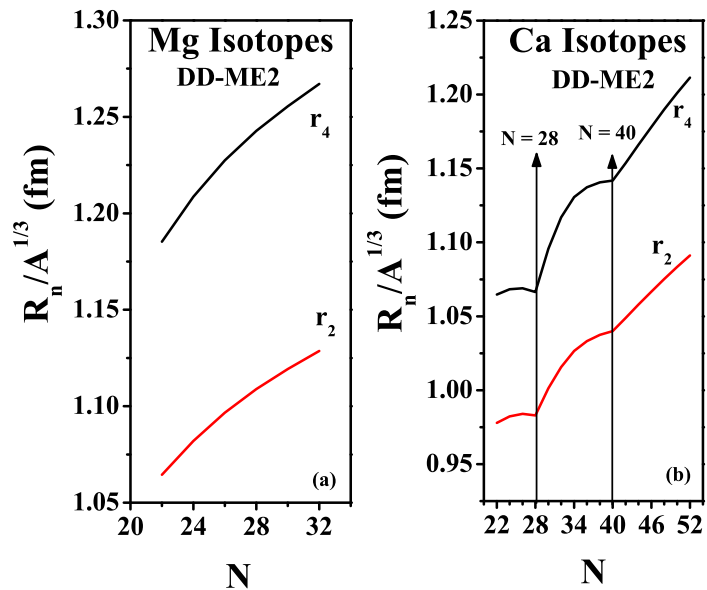}
\caption{(Colour online) Second and fourth radial moments r$_{2}$ and r$_{4}$ \cite{reinhard2020} for Mg and Ca isotopes.}
\label{fig4}
\end{figure}

Fig. \ref{fig4} shows the rms charge radius (second moment r$_{2}$) and fourth radial moment (r$_{4}$) plotted for Mg and Ca isotopes. For Ca isotopes, a well pronounced kink is observed at N $=$ 28 and 40 in both the curves of second and fourth radial moment r$_{2}$ and r$_{4}$ which demonstrates the strong magicity at $^{48}$Ca and $^{60}$Ca. In case of Mg isotopes, the absence of any kind of bend or kink in neutron-rich Mg at N$=$28 in both r$_{2}$ and r$_{4}$ curves supports the collapse of magicity in $^{40}$Mg. However the presence of a kink in $^{48}$Ca ensures the magicity of N$=$28 in stable nucleus $^{48}$Ca. It also shows the emergence of a new magic number N$=$40 in $^{60}$Ca near neutron drip-line. \par

\begin{figure}[htb]
\centering
\includegraphics[width=0.8\textwidth]{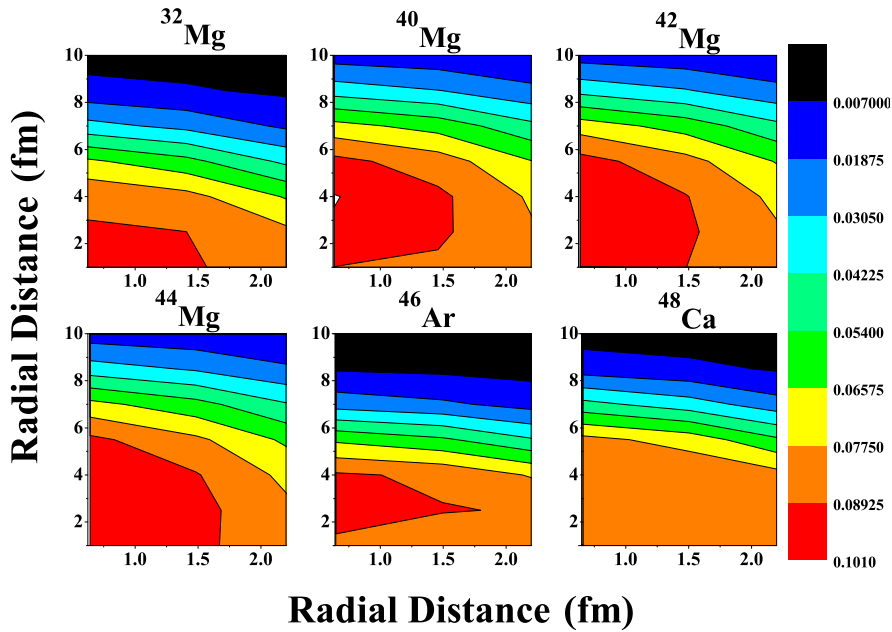}
\caption{(Colour online) Radial density distribution of N$=$20 and 28 isotones using RMF(NL3*) approach.}
\label{fig5}
\end{figure}

\subsection{Weakly bound halo like structure and 2n-decay in Mg isotopes near drip-line}

Nuclear halos appear when the valence neutrons have zero or very low orbital angular momentum. In case of $^{40}$Mg, the occupancy of valence neutrons in higher orbital 2p$_{3/2}$ in pf-shell with low orbital angular momentum ($\ell$$=$1) than the state 1f$_{7/2}$ ($\ell$$=$3) (as seen in Fig. \ref{fig1}) results in the smaller centrifugal barrier which leads to the wider density distribution indicating a halo. Since this nucleus is highly deformed, the occupancy in Nilsson orbital with mixing of f-p states appears to be responsible for weakly bound halo structure.  Our calculations show the occupancy of approximately 6 particles in f$_{7/2}$ (5/2-[312], 3/2-[321], 1/2-[330]) state and 2 valence neutrons in p-states (1/2-[310]) which contribute to the pairing interaction, weak binding and 2n-halo. \par

Fig. \ref{fig5} shows the radial density distribution for few neutron-rich isotones of N$=$20 and 28. A long tail of well extended density distribution in $^{40}$Mg  indicates a halo. Also, a similar density tail in $^{42,44}$Mg shows the existence of a deformed halo as also indicated in Refs.$~$\cite{LLI,zhou,meng2015} for $^{42,44}$Mg. The density distribution in $^{40,42,44}$Mg is much more extended than in the other nuclei of similar mass range considered here. This makes $^{40,42,44}$Mg isotopes very strong candidates of deformed halo. However, since $^{42,44}$Mg are found to be 2n-unbound as seen in Table I, it appears to be the large deformation effect that increases the number of neutrons in the classically forbidden regions below the continuum threshold along with the scattering of valence neutrons due to pairing correlation, that are contributing to halo like structure as well as an unbound state with a very little lifetime which will be discussed later in this section. \par

\begin{figure}[htb]
\centering
\includegraphics[width=0.6\textwidth]{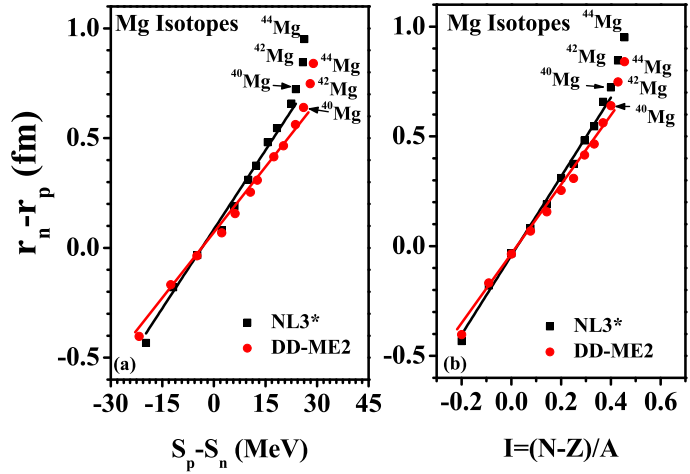}
\caption{(Colour online) Neutron skin thickness of Mg isotopes vs. $S_{p} - S_{n}$ and $I=(N-Z)/A$.}
\label{fig6}
\end{figure}

Due to extended density distribution of loosely bound neutrons, the radii is expected to grow larger which is a typical signature of the presence of large skin or halo structure in nuclei ~\cite{tanihata,tanihata1,watanabe2014}. In Fig. \ref{fig6}, we plot neutron-skin thickness (r$_{n}$$-$r$_{p}$)  vs. the difference of separation energies ($S_{p} - S_{n}$) (Fig. \ref{fig6}a) \cite{suzukiNa} and neutron excess $I=(N-Z)/A$ (Fig. \ref{fig6}b) \cite{warda} of Mg isotopes. The neutron skin shows a linear dependence on $S_{p} - S_{n}$ and $I=(N-Z)/A$, which was seen in similar investigations \cite{suzukiNa,warda} reported for Na isotopes. But as we approach neutron-rich $^{40,42,44}$Mg, the sudden rise in the neutron skin thickness with change of slope and the departure from linear curve demonstrates the halo like structure for these nuclei. \par

\begin{figure}[htb]
\centering
\includegraphics[width=0.7\textwidth]{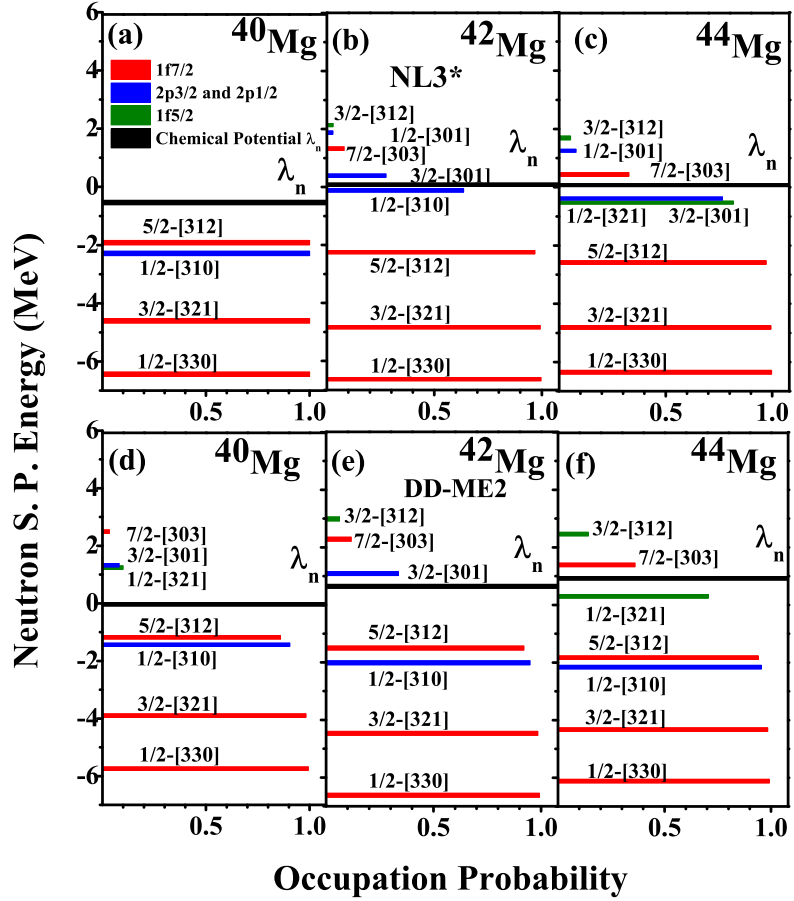}
\caption{(Colour online) Neutron s.p states of pf-shell vs. occupation probability for $^{40,42,44}$Mg.}
\label{fig7}
\end{figure}

A valence 2n-weakly bound system that appears to form a halo like structure in the low density nuclear matter far from the tightly bound core may lead to the enhanced probability of direct 2n-decay as appears in $^{42,44}$Mg. This characteristic where the two-neutron radioactivity is correlated to the halo like structure shown in this work, has not been shown in any other work as far to our knowledge. Direct 2n-decay from neutron-rich nuclei is an analogous process of 2p-decay \cite{Saxena2017,hagino2014}, characterized by the tunneling of two neutrons through a centrifugal barrier. Due to absence of long range Coulomb interaction in 2n-radioactivity, the dineutron correlation are relatively easy to probe using 2n-decays for which $^{42,44}$Mg may be potential candidates. At the same time, due to the absence of Coulomb barrier, the shorter lifetimes of the neutron decaying states pose a serious experimental challenge in the probe of two-neutron unbound system \cite{kohley,grigorenko} especially for higher masses. Hence looking at the practical difficulties in probing 2n-unbound systems, one has to rely on the theoretical models and a comprehensive study of such 2n-unbound systems.  \par

The scattering of particles from bound states to continuum contributes to the pairing interaction which influences the decay modes of halo nuclei. Thus, we study neutron s.p. spectrum of pf-shell for $^{40,42,44}$Mg using NL3* (Fig. \ref{fig7} (a),(b),(c)) and DD-ME2 (Fig. \ref{fig7} (d),(e),(f)) parameters which show excellent agreement. The orbital mixing of f-p Nilsson states and occupancy of p-states (1/2-[310], 3/2-[301], 1/2-[301]) contribute more to pairing interaction. As seen in Fig.2, $^{40}$Mg has more pairing energy and hence more binding than $^{42,44}$Mg that  prohibits 2n-simultaneous emission in  $^{40}$Mg and keeps it weakly bound. Additional neutrons in $^{42,44}$Mg seems to have reduced the pairing energy and binding which support 2n-emission. It also hints that the 2n-emitters, in general, may have halo like structure as in $^{42,44}$Mg but the existence of halo may not necessarily lead to the 2n-radioactivity as in $^{40}$Mg.\par

One interesting and important  observation in Fig. \ref{fig7} is the 1f$_{5/2}$ state, which is essentially a positive energy state lying at the continuum threshold, has very small  but finite occupancy in $^{42}$Mg and $^{44}$Mg. The presence of f-states for the valance neutrons (Fig. \ref{fig7}) in $^{42,44}$Mg may enable the valence neutrons to get confine to the nucleus due to larger centrifugal barrier enabling the nucleus to acquire some more life time which shows agreement with the prediction of $^{42}$Mg  to be a 2n-emitter and a long lived nucleus \cite{thoennessen2004,pei2009} as well. This additional stability may delay the 2n-emission in $^{42,44}$Mg and provide some life time sufficient for experimental probes, but may not be adequately sufficient to hold the nucleus bound. This makes $^{42,44}$Mg a promising candidates of deformed halo that favour 2n-radioactivity and may provide some life time to probe dineutron correlations using 2n-decay. \par

Here it is important to point out that since the Coulomb barrier is absent in 2n-decay, the two-neutron decay has an advantage over the two-proton decay that even for the smallest decay energies the lifetimes will be small compared to the $\beta$-decay lifetimes \cite{thoennessen2004} and hence the 2n-decay would be a favoured decay mode. However, the experimental difficulties related to the lesser lifetime due to the absence of Coulomb barrier pose another challenge to deal with the presently available techniques which gives impetus to the theoretical investigation of such 2n unbound systems. Our RMF calculations with NL3* parameter estimated the decay energies of 60 keV and 270 keV for $^{42}$Mg and $^{44}$Mg, respectively which are very small. With these decay energies, the half-lives of two neutron decay can be estimated which may be a significant milestone in the probe of 2n-decay but that would be possible only with the availability of suitable experimental techniques. This needs attention and extensive investigation. Some of our preliminary theoretical calculations of 2n-decay lifetimes have provided some insights but it needs much more systematic and detailed investigation  which may be reported soon in our upcoming works.\par
\section{Conclusion}
To conclude, we present a systematic study of the neutron-rich experimentally accessible last bound Mg isotope $^{40}$Mg,  and 2n-unbound $^{42,44}$Mg. 2n-halo and 2n-radioactivity is probed by employing the RMF approach using various parameters/variants. The extended spatial density distributions, enhanced neutron radius and skin thickness, pairing correlations, and wave functions predict $^{40,42,44}$Mg to be deformed neutron halo systems which enhances the possibility of 2n-radioactivity. Weakening of magicity at N$=$28 plays a major role in the existence of a halo in $^{40}$Mg. A strong correlation between halo and the collapse of magicity of N$=$28, deformation, mixing of f-p shell orbitals and pairing interactions has been shown. Delayed emission  due to centrifugal barrier and pairing correlations make $^{42,44}$Mg potential candidates for probing dineutron correlations.\par
\section{Acknowledgements}
G.S. and M.A. thank SERB (DST) for support under CRG/2019/001851 and WOS-A scheme, respectively.

\end{document}